\documentstyle[12pt,a4,epsfig,a4wide]{article} 

\def\Jo#1#2#3#4{{#1} {\bf #2}, #3 (#4)}

\def\NPB{{\em Nucl. Phys.} {\bf B}}

\def\PLB{{\em Phys. Lett.}  {\bf B}}
\def\PRL{\em Phys. Rev. Lett.}
\def\PRD{{\em Phys. Rev.} {\bf D}}

\def\EPC{{\em Eur. Phys. J.} {\bf C}}

\def\ra{\rightarrow}
\def\be{\begin{equation}}
\def\ee{\end{equation}}

\def\gs{\mathrel{
   \rlap{\raise 0.511ex \hbox{$>$}}{\lower 0.511ex \hbox{$\sim$}}}}
\def\ls{\mathrel{
   \rlap{\raise 0.511ex \hbox{$<$}}{\lower 0.511ex \hbox{$\sim$}}}}
\def\obb{0\mbox{$\nu\beta\beta$}}
\def\onbb{neutrinoless double beta decay}
\newcommand{\Slash}[1]{\mbox{$#1\hspace{-.6em}/$}}
\def\ba{\begin{array}{c}}
\def\baz{\begin{array}{cc}}
\def\bad{\begin{array}{ccc}}
\def\bea{\begin{equation} \begin{array}{c}}
\def\eea{ \end{array} \end{equation}}
\def\ea{\end{array}}
\def\D{\displaystyle}

\def\aen{\mbox{$\overline{\nu_e}$ }}

\begin{document}
\newpage
\title{
\hfill { \bf {\small DO--TH 00/11}}\\
\hfill { \bf {\small hep-ph/0005270}}\\ \vskip 1cm  
\bf 
Signatures of heavy Majorana neutrinos and HERA's isolated 
lepton events}
\author{W. Rodejohann$^a$\footnote{Email address:
rodejoha@dilbert.physik.uni-dortmund.de}$\;$ , 
K. Zuber$^b$\footnote{Email address:
zuber@physik.uni-dortmund.de}\\ \\
{\it $^a$Lehrstuhl f\"ur Theoretische Physik III,}\\
{\it $^b$Lehrstuhl f\"ur Experimentelle Physik IV,}\\ 
{\it Universit\"at Dortmund, Otto--Hahn Str.4,}\\
{\it 44221 Dortmund, Germany}}
\date{}
\maketitle
\thispagestyle{empty}
\begin{abstract}
The graph of neutrinoless double beta decay is 
applied to HERA and generalized to final states with any two 
charged leptons. Considered is the case in which 
one of the two escapes typical 
identification criteria and the case when 
a produced tau decays hadronically. 
Both possibilities 
give one isolated lepton with high transverse momentum, hadronic activity and 
an imbalance in transverse momentum. 
We examine the kinematical properties of these events and compare them 
with the high $p_T$ isolated leptons reported by the H1 
collaboration. Their positive charged muon events can 
be explained by the ``double beta'' 
process and we discuss possibilities for the precise determination 
which original final state produced the single isolated lepton. 
To confirm our hypothesis one should search in the data for 
high pseudorapidity and/or low $p_T$ leptons or for additional separated 
jets.    
\end{abstract}

\newpage
\section{\label{intro}Introduction}
Despite the impressive confirmation of predictions from 
the standard model (SM) it is general believe that we are on the verge 
of fundamental new discoveries, be it production of new 
particles or significant 
deviations of observables in high--precision measurements. 
Effects of new physics might also be hidden in existing data sets 
and it is interesting to see what candidates are able to explain 
any unexpected events or measurements. 
A first step in this direction was done in terms of 
observation of nonvanishing neutrino rest masses, most clearly seen in 
the up--down asymmetry of the atmospheric muon neutrino flux  
in SuperKamiokande \cite{SK}. 
The smallness of these masses can be related to massive new particles via 
the see--saw--mechanism \cite{seesaw}. 
In this respect it seems most natural to look for effects of 
massive neutrinos, i.e.\ search for hints of these 
new particles in high-- or 
low--energy experiments. The theoretical prejudice is that 
the neutrinos are Majorana particles --- be it because they are 
delivered by see--saw or pop out of almost every GUT --- and 
we shall follow this idea.\\ 
Especially for the case of heavy (few 100 GeV) Majorana neutrinos,  
production at accelerators has been investigated by 
many authors. 
The different possibilities include $e^+ e^-$ \cite{epem}, 
$pp$ \cite{pp}, $p \overline{p}$ \cite{ppquer}, $\nu N$ \cite{FRZ1}, 
$ep$ \cite{ep,FRZ2}, 
linear colliders and $e^- e^-$ \cite{ee1,ee2} or even $e \mu$ machines 
\cite{emu}.\\
Heavy Majoranas have also been studied within the context of 
low--energy experiments such as neutrinoless double 
beta decay (\obb{}) \cite{ee1,2b} or Kaon decays \cite{kaon}.  
The respective Feynman diagrams as well as the concrete model differ in 
most publications and the interested reader might compare the 
papers with respect to that.\\ 
One of the anomalies in existing data is the existence of 
high $p_T$ isolated leptons together with large missing 
transverse momentum ($\Slash{p}_T$) at HERA\@. Since the first 
event \cite{isol1} was discovered by H1, five more were found 
\cite{isol2} and at least 3 of them can not be explained by 
$W$ production or other 
SM processes. In contrast to that, ZEUS sees no excess in these 
events \cite{isozeus}, yet, at the present statistical level, 
there is no contradiction \cite{cozzika}.\\
In \cite{FRZ2} we examined the process (see Fig.\ \ref{diagramm}) 
\be \label{process}
e^+ p \ra \aen \alpha^+ \beta^+ X \mbox{ with } 
\alpha , \, \beta = e , \, \mu , \, \tau 
\ee 
and discussed possible signals of this like--sign dileptons (LSD) and 
high $\Slash{p}_T$ final state. 
No such events are reported and previously unavailable direct 
limits on the elements of the Majorana mass matrix were derived 
\cite{FRZ2,ich}. 
However, it turns out that when the kinematical cuts used in H1's search 
for isolated leptons are applied to our process (\ref{process}), they  
tend to ignore one of the two leptons.  
Especially the requirement of 
$p_T^{\rm lepton} > 10$ GeV is often too much for both charged leptons to 
fulfill. 
The LSD signal of Eq.\ (\ref{process}) is thus 
reduced to {\it one} isolated lepton with high $\Slash{p}_T$. 
This possibility can be checked by looking for an additional isolated 
low $p_T$ and/or high pseudorapidity lepton.  
In addition, it is possible that a produced $\tau$ decays 
hadronically\footnote{We shall use the term electron, muon or tau 
for both, particle and antiparticle.}, 
resulting also in single lepton final states. 
More than one isolated jet would be a signal for this kind of event. 
Since process (\ref{process}) gives LSD with the same sign as 
the incoming lepton we concentrate on H1's positive charged 
muon events, since there are no positron events found. This 
fact might be explained if one incorporates also limits on mixing 
of heavy neutrinos, as derived from \obb. With this constraint the 
expected $e$ signal 
is smaller than the $\mu$ signal.\\ 
One might argue that direct Majorana ($N$) production via a $e^+ N W$ 
vertex is more likely to occur since the cross section is larger. 
At present there is only an analysis in HERA's $e^- p$ mode available 
\cite{ep} and it was found that detection is only possible if the $N$ decays 
into $e^+ W^-$, giving an isolated lepton with different charge than 
the incoming one. The reason for that is of course the large 
background from $W$ production. However, a general analysis of all 
channels ($N \to \nu Z$, $N \to \mu W , \, \ldots$) 
remains still to be done and it might be interesting to compare the 
results with our signals in the future. 
Until that is done, we think that our process is worth considering, 
inasmuch as we have no restriction to the flavor of the final 
state lepton.\\
The paper is organized as follows: 
In Section \ref{zwei} we discuss some general features of the process and 
the diagram and argue in Section \ref{2to1} 
that the two lepton signal of Eq.\ (\ref{process}) might very 
well be seen as a one lepton signal.  
Section \ref{sigobs} sees a discussion of signals of the events and how one 
might distinguish the genuine final state from the measured one. 
Finally, Section \ref{concl} closes the paper with 
a conclusion and discussion.

\section{\label{zwei}The process and heavy Majorana neutrinos}
\begin{figure}[hb]
\setlength{\unitlength}{1cm}
\epsfig{file=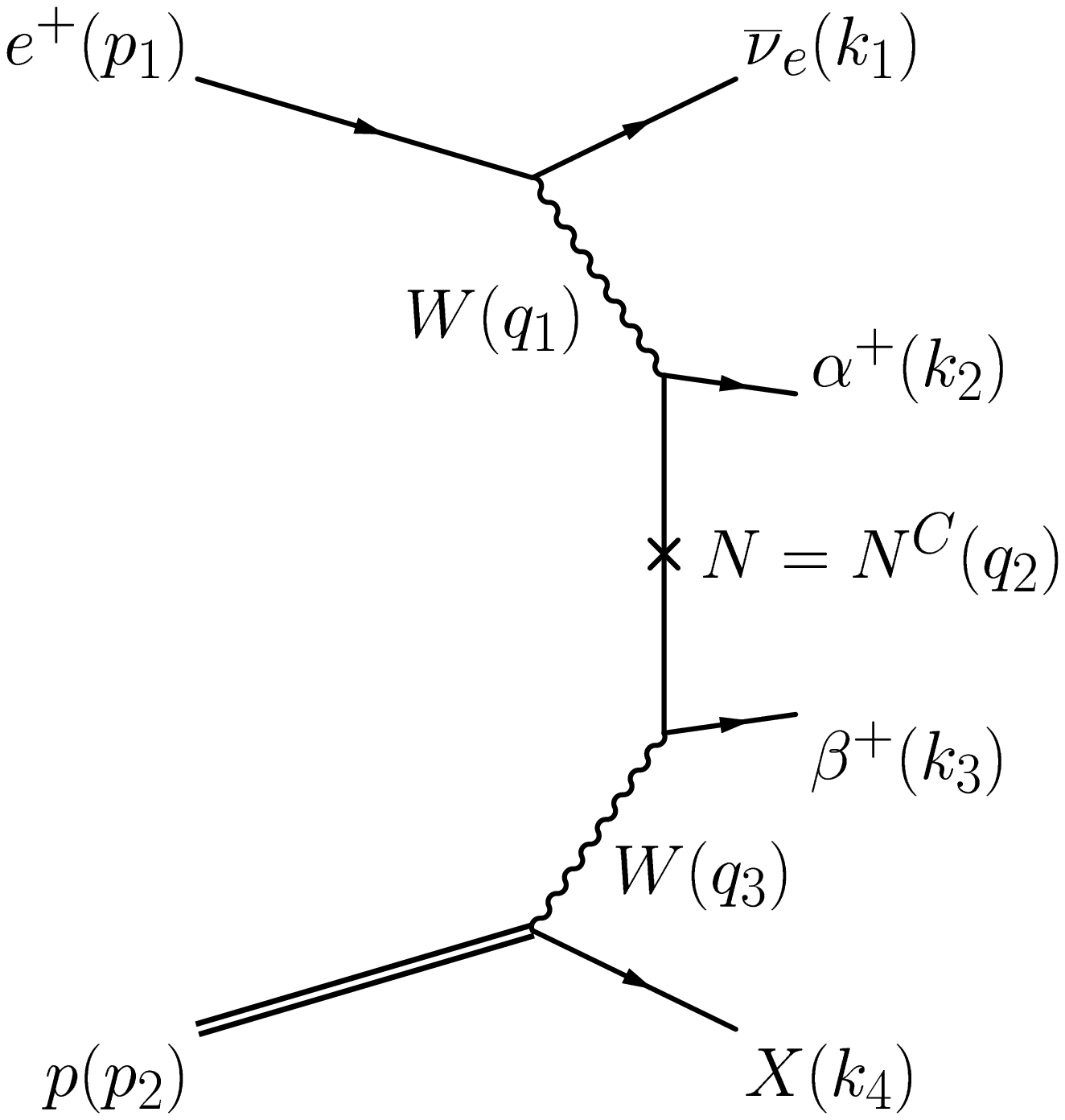,width=14cm,height=15cm}
\vspace{-5cm}
\caption{\label{diagramm}Diagram for 
$e^+ p \rightarrow \overline{\nu}_e \alpha^+ \beta^+ X$. 
Note that there is a crossed term and for $\alpha \neq \beta$ there are 
two possibilities for the leptons to be emitted from.}
\end{figure}

\begin{figure}[hb]
\setlength{\unitlength}{1cm}
\vspace{-1cm}
\hspace{-1.4cm}
\epsfig{file=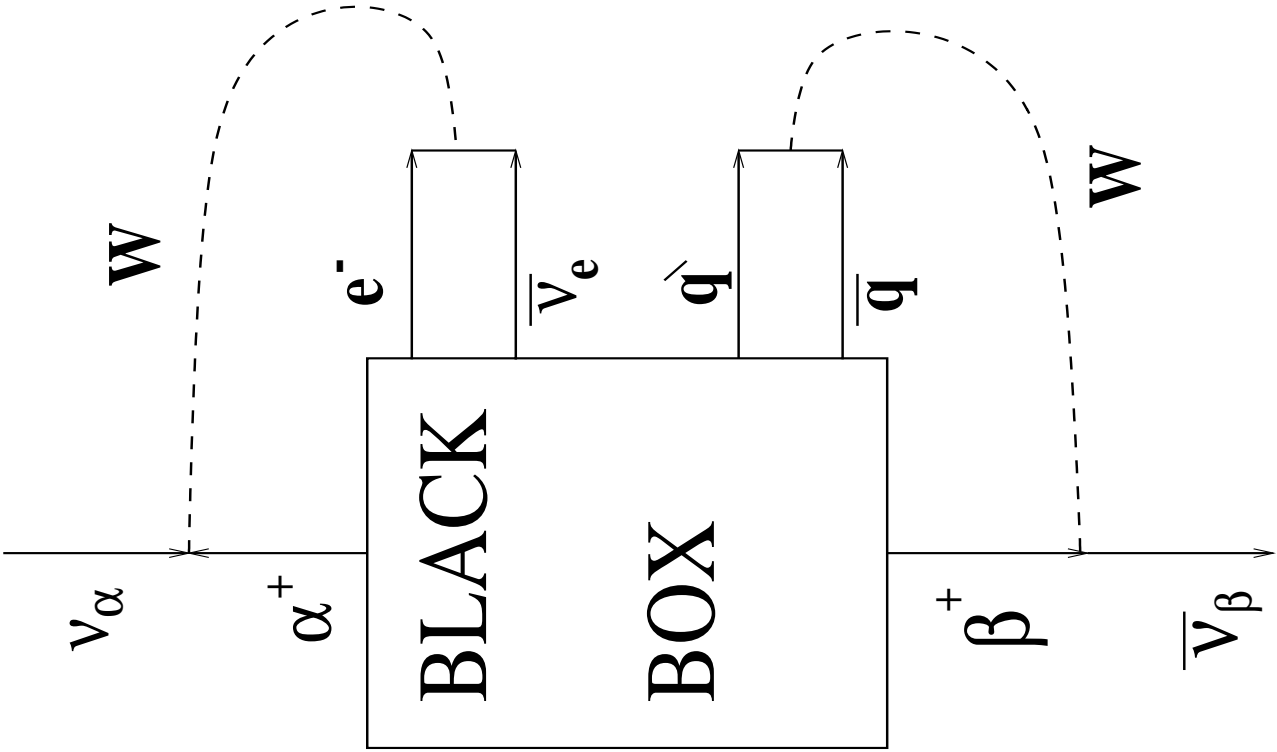,width=17cm,height=10cm}
\caption{\label{blackbox}Connection between Majorana mass term of 
$\nu_\alpha$ and $\nu_\beta$ and the existence of process (\ref{process}).}
\end{figure}

We shall work in a mild extension of the SM with no further 
specification of how heavy Majoranas might be created. The coupling 
to the usual leptons and gauge bosons is the familiar left--handed 
weak interaction. 
The three known light neutrinos $\nu_\alpha$ 
are thus mixtures of light and heavy mass particles, this can be 
expressed by the replacement 
\be
\nu_\alpha \to \cos \theta_\alpha \nu_\alpha 
+ \sin \theta_\alpha N_\alpha
\ee
in the (unmixed) Lagrangian for each family. 
For the sake of simplicity we take $N_\alpha = N$. 
The Lagrangian now reads: 
\bea \label{lag}
-\mbox{${\cal L}$} = \frac{\D g}{\D \sqrt{2}} W_\mu \left\{ \cos \theta_\alpha 
\overline{\nu_\alpha} \gamma^\mu \gamma_- l_\alpha 
+ \sin \theta_\alpha \overline{N} 
\gamma^\mu \gamma_- l_\alpha \right\} \\[0.3cm]
+  \frac{\D g}{\D 2 \cos \theta_W} Z_\mu 
\left\{ \cos^2 \theta_\alpha \overline{\nu_\alpha}  
\gamma^\mu \gamma_- \nu_\alpha + 
\sin 2 \theta_\alpha \overline{N} \gamma^\mu \gamma_- \nu_\alpha - 
\frac{\D 1}{\D 2} \sin^2 \theta_\alpha  \overline{N}  
\gamma_\mu \gamma_5 N \right\}
+ \rm h. c.  
\eea
where $\gamma_- = \frac{1}{2} (1 - \gamma_5$) and 
there is no vector current between $\overline{N}$ and $N$ due to their 
Majorana nature. We can keep it for the light neutrinos since for 
energies much larger than the (light) masses there is hardly a chance to 
find a difference between $2 \overline{\nu} \gamma^\mu \gamma_- \nu$ 
and $\overline{\nu} \gamma^\mu \gamma_5 \nu$  
\cite{kayser}.\\ 
What can we expect for the values 
of the masses and the mixing parameters? 
Taking the typical see--saw formula 
we find 
\be
m_\nu \simeq \frac{\D m_D^2}{\D m_N} \Rightarrow m_N \simeq 
\frac{\D (10^5 \ldots 10^{11})^2}{\D 10^{-5} \ldots 1} \rm \; eV  
\simeq 100 \ldots 10^{18} \; GeV
\ee
where we took for the Dirac mass $m_D$ every value from electron to 
top mass and for the light neutrino mass we allowed everything from the 
vacuum solution in a highly hierarchical scheme 
($\sqrt{\Delta m^2 } \simeq m_\nu \simeq 10^{-5}$ eV) 
to a degenerate scheme (cosmological or also LSND's mass scale) 
$\sum m_\nu \simeq $ few eV (see \cite{ich} for a detailed analysis 
of allowed schemes). For the mixing angle we have 
\be 
\theta_\alpha \simeq \sin \theta_\alpha = \frac{\D m_D}{\D M_N}  \simeq 
\frac{\D m_\nu}{\D m_D} \simeq 10^{-5} \ldots 10^{-16}  . 
\ee
However, we shall use the current bounds on $\theta_\alpha$ 
which are \cite{mixbound} 
\be \label{mix} 
\sin^2 \theta_e \le 6.6 \cdot 10^{-3},  \;
\sin^2 \theta_\mu \le 6.0 \cdot 10^{-3} \mbox { and } 
\sin^2 \theta_\tau \le 1.8 \cdot 10^{-2} \, . 
\ee
Note that the lowest value is for the muon sector. Eq.\ (\ref{lag}) 
can now be applied to calculate the width of the Majorana, which 
is dominated by the two--body 
decays $N \to W \alpha$ and $N \to Z \nu_\alpha$, 
we find 
\bea
\Gamma (N) = \D\sum\limits_\alpha 
\frac{\D G_F \sin^2 \theta_\alpha}{\D 8 \pi \sqrt{2} M_N^3}
\left\{ 
\left[ (M_N^4 - M_W^4) + M_W^2 ( M_N^2 - M_W^2) \right] (M_N^2 - M_W^2) 
\right. \\[0.4cm]
\D \left. 
+ \cos^2 \theta_\alpha 
\left[ (M_N^4 - M_Z^4) + M_Z^2 ( M_N^2 - M_Z^2) \right] (M_N^2 - M_Z^2) 
\right\} . 
\eea
Direct searches for heavy neutrinos give typical lower limits 
\cite{massbound} on their mass of 70 to 100 GeV, depending on their character 
(Dirac in general gives a higher bound) and to which lepton 
family they couple to. Unfortunately, the maximal value of the cross 
section of process (\ref{process}) in Fig.\ \ref{diagramm} 
is found to lie in that range as well \cite{FRZ1,FRZ2}. The 
dependence on the mass goes as 
\be
d \sigma \propto \frac{\D M_N^2}{\D (q^2 - M_N^2)^2} 
\rightarrow \left\{ \baz M_N^2 & \mbox{ for } M_N^2 \ll q^2 \\[0.3cm]
                         M_N^{-2} & \mbox{ for } M_N^2 \gg q^2 \ea \right. , 
\ee
where $q$ is the momentum of the Majorana. The standard calculation 
gives for the matrix element \cite{FRZ1} (see Fig. \ref{diagramm} 
for the attachment of momenta): 
\bea
|\overline{\mbox{$\cal{M}$}}|^2 (e^+ q \to \aen \alpha^+ \alpha^+ q') 
= \sin^4 \theta_\alpha \, M_N^2 \, G_F^4 \, M_W^8 \,  
2^{12} \, \frac{\D 1 }{\D (q_1^2 - M_W^2 )^2(q_3^2 - M_W^2)^2} 
(k_1 \cdot p_2)  \\[0.4cm]
\left[ \frac{\D 1}{\D (q_2^2 - M_N^2)^2} (k_2 \cdot p_1)(k_3 \cdot k_4) + 
\frac{\D 1}{\D (\tilde{q_2}^2 - M_N^2)^2}(k_3 \cdot p_1)(k_2 \cdot k_4) 
\right. \\[0.4cm]
\left. - \frac{\D 1}{\D (q_2^2 - M_N^2)(\tilde{q_2}^2 - M_N^2)} 
\mbox{\Large(} 
(k_2 \cdot k_3)(p_1 \cdot k_4) - (k_2 \cdot p_1)(k_3 \cdot k_4) 
- (k_3 \cdot p_1)(k_2 \cdot k_4) \mbox{\Large)} \right]
\eea
and the scattering with an antiquark sees $k_4$ interchanged with $p_2$. 
Here $\tilde{q_2}$ denotes the momentum of the Majorana in the 
crossed diagram, which has a relative sign due to the interchange 
of two identical fermion lines. 
In addition one has to include a factor $\frac{1}{2}$ to avoid 
double counting in the phase space integration. For the phase space 
we called the routine GENBOD \cite{genbod} and for the parton 
distributions we applied GRV 98 \cite{grv98}. In case a $\tau$ is produced 
we additionally folded in its three--body decay.
We inserted finite ($W$ and $N$) width effects in our program 
and found them to be negligible.\\
An interesting statistical effect occurs when one considers the relative 
difference between, say, the $\mu\mu$ and the $\mu \tau$ final state 
(mass effects play no significant role): 
First, there is no factor $\frac{1}{2}$ for the latter case. Then, there 
is the possibility that a $\tau$ is produced at the (``upper'') 
$e^+ \aen W$ vertex or at the (``lower'') $q q' W$ vertex. Both diagrams are 
topologically distinct and thus have to be treated separately. 
This means, four diagrams lead to the $\mu\tau$ final state, whereas only 
two lead to the $\mu\mu$ final state. 
We see that there is a relative factor 4 between the two cases. 
Note though that now the interference terms are {\it added} 
to the two squared amplitudes since there is no relative 
sign between the two. This reduces the relative factor to about 3. 
However, effects of kinematical cuts and the limits of Eq.\ (\ref{mix}) wash 
out this phenomenon. 
A similar situation occurs when one studies the $\tau\tau$ case and 
lets the $\tau$'s decay into different particles (e.g.\ $e\nu\nu$ and 
$\mu \nu \nu$). There is no way to tell into what the 
``upper'' or ``lower'' tau decays, so one has to include both cases.\\
A question arises if one can conclude a Majorana mass term if we measure a 
process like Eq.\ (\ref{process}). Here, a simple generalization of 
arguments first given by Schechter and Valle \cite{scheva} 
for \onbb{} applies: 
Assuming we found indubitable evidence for 
$e^+ q \to \aen{} \alpha^+ \beta^+ q'$, then crossing permits the process 
$0 \to e^- \aen{}  \; \alpha^+ \beta^+ \; q' \overline{q}$, realized 
by the ``black box'' in Fig.\ \ref{blackbox}. 
Any reasonable gauge theory will have $W$'s couple to quarks and 
leptons, so that a Majorana mass term 
for $\nu_\alpha$ and $\nu_\beta$ is 
produced by coupling one $W$ to the $\alpha^+ e^- \aen$ and one $W$ 
to the $\beta^+ q' \overline{q}$ vertex. 
Since we do not know which two quarks participate and which 
neutrino couples to the positron (the Schechter and Valle argument 
for \obb{} works with two pairs of $u$ and $d$ quarks), 
this theorem 
holds for a greater class of models, namely e.g.\ those with direct 
$e^- \stackrel{(-)}{\nu}_{\! \! \!X}$ coupling, with $X$ being any flavor.\\
The connection between a \onbb{} signal and Majorana masses 
has been expanded in \cite{klapsusy} to supersymmetric theories 
and it was found that it implies Majorana masses also for sneutrinos, 
the scalar superpartners of the neutrinos.\\
However, in contrast to the signal in \onbb{} experiments 
(two electrons with constant sum in energy) the identification of 
our process will be very difficult and deciding which $\alpha\beta$ 
final state was originally produced remains a hard task. In addition, 
the number of expected events turns out to be far less than one. 
Nevertheless, the demonstration 
of Majorana mass terms will be an exciting and important result, since 
different models predict different texture zeros in the mass matrix. 
In some models the $ee$ entry in the mass matrix 
is zero and therefore the only direct information 
about the mass matrix might come from neutrino oscillations, 
cosmological considerations, global fits 
and direct searches, e.g.\ at LEP\@.   
This complicates the situation, since e.g.\ in oscillations 
only mass squared differences are measured and the additional 
phases induced by the Majorana nature are unobservable. To combine all 
information from the different approaches input from 
models is required. 
Thus, the exact form of the matrix is highly nontrivial to find. 
Therefore, information about non--vanishing 
entries in the mass matrix is very important and one has to take every 
opportunity to find out about all elements and the Majorana character 
in general. 
In addition, if such a lepton--number violating process is detected, 
it is surely helpful to 
know if the ``mildly extended'' SM can provide the signal or 
another theory, such as SUSY, has to be blamed.

\section{\label{2to1}Two become one}
\begin{figure}[hb]
\setlength{\unitlength}{1cm}
\vspace{0.3cm}
\begin{center}
\epsfig{file=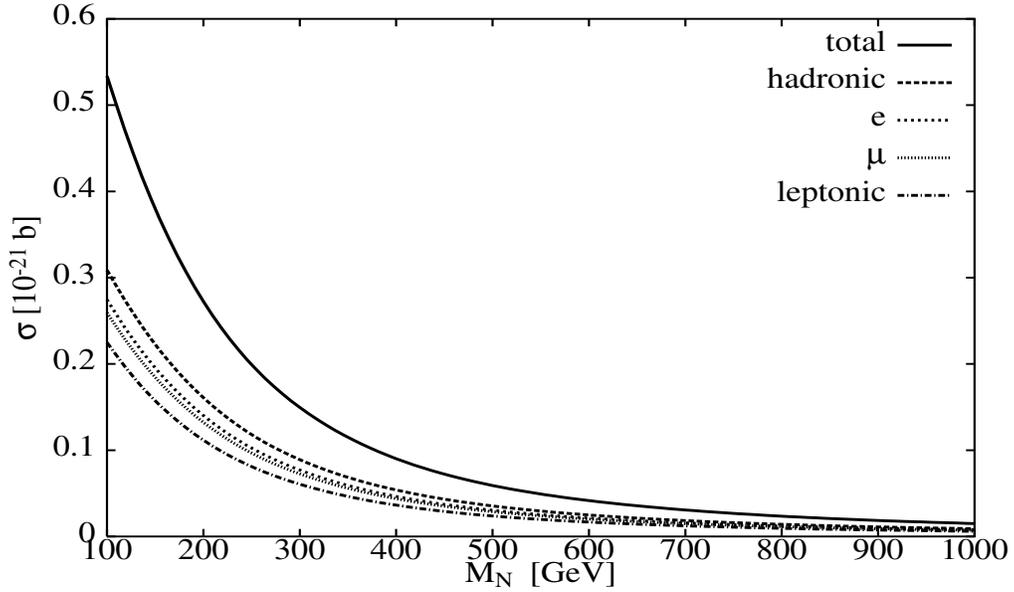,width=13cm,height=8cm}
\end{center}
\caption{\label{sigvonm}Cross section of the 
expected isolated lepton signal 
from all possible final states as a function of the Majorana mass $m_N$.} 
\end{figure}

\begin{figure}[hb]
\setlength{\unitlength}{1cm}
\vspace{0.3cm}
\begin{center}
\epsfig{file=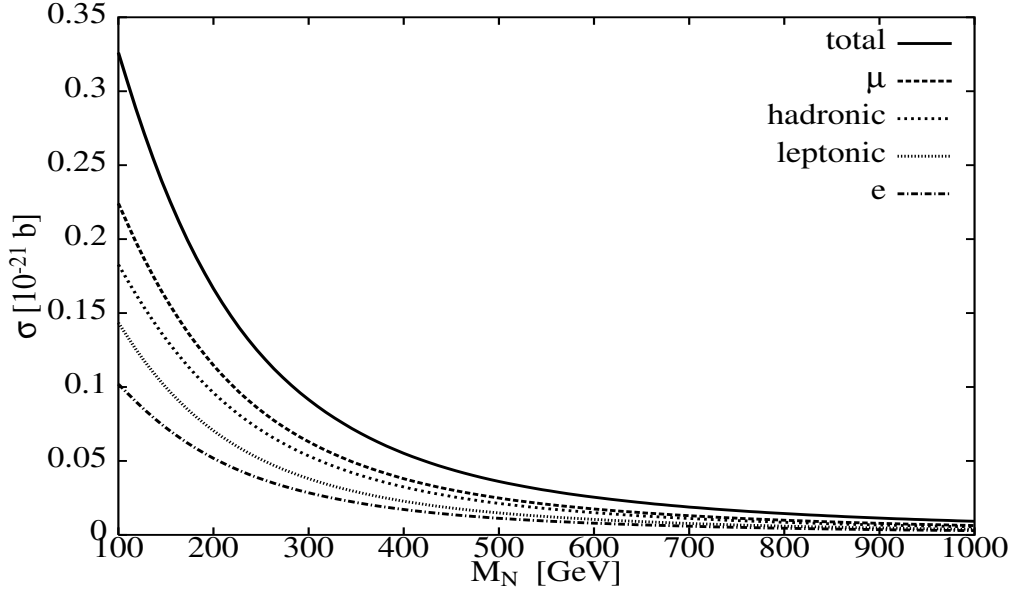,width=13cm,height=8cm}
\end{center}
\caption{\label{sigvonm1}Same as previous figure if one also incorporates 
the \obb{} limit from Eq.\ (\ref{obbl}).} 
\end{figure}

\begin{figure}[hb]
\setlength{\unitlength}{1cm}
\vspace{0.3cm}
\begin{center}
\epsfig{file=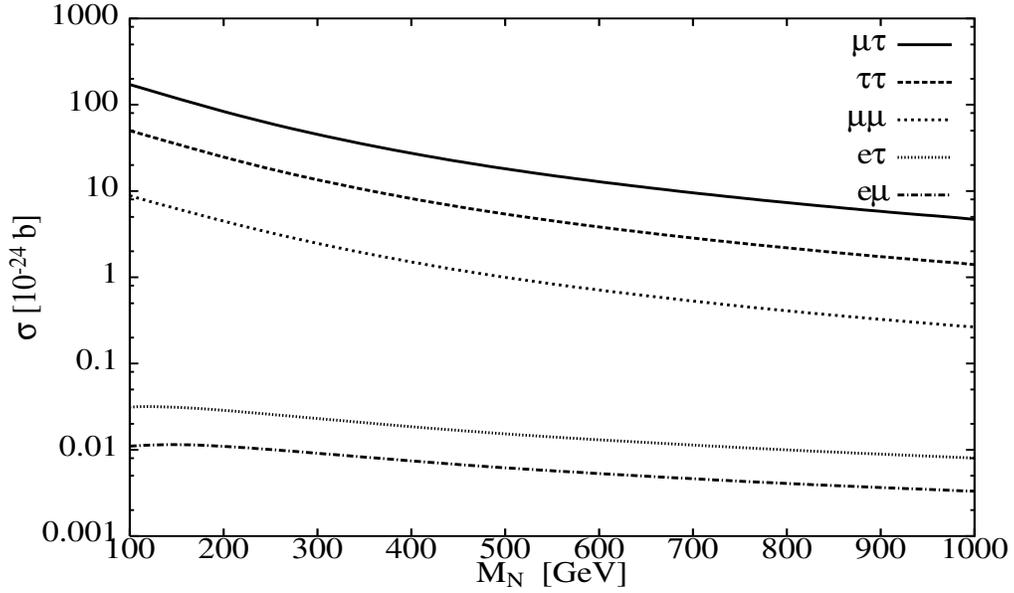,width=13cm,height=8cm}
\end{center}
\caption{\label{muonsigvonm}Contribution of the different final 
states to the cross section of the muon signal as a 
function of the Majorana mass $m_N$.} 
\end{figure}

We applied the same cuts as H1 in their search for isolated leptons:
\begin{itemize}
\item 
Imbalance in transverse momentum $\Slash{p}_T \ge 25$ GeV
\item 
Transverse momentum of lepton $p_T \ge 10$ GeV
\item
Pseudorapidity of lepton $|\eta| \le 2.436$  
\item
Distance between charged lepton and closest jet 
in $\eta$--$\phi$ space\footnote{Actually H1's value is 
1.0 or 0.5, depending on the way they define jets for their 
respective analysis. We use a 
general value of 1.5 to account for hadronization effects.}, 
$\Delta R \ge 1.5$ where $\phi$ is the azimuthal angle 
\item
Angle of the hadronic jet(s) $4^\circ \le \theta^X \le 178^\circ$
\end{itemize}
This has to be compared with our cuts in \cite{FRZ2}, 
$\Slash{p}_T \ge 10$ GeV, $|\eta| \le 2.0$ for all 
measured particles and $\Delta R \ge 0.5$ between the charged leptons and the 
hadronic remnants. It turns out that both sets of cuts deliver cross sections 
in the same ballpark. 
There are now two possibilities for the original LSD signal to 
appear as one single lepton:
\begin{itemize}
\item[1.]One lepton can have high pseudorapidity and/or low $p_T$. 
This lepton then also contributes to the missing transverse momentum.
\item[2.]If one tau is produced it might decay hadronically, adding 
one neutrino to the imbalance in $p_T$ and also additional hadronic jets.
\end{itemize} 
A detailed analysis of the LSD signal might be done if one finds such 
events. 
Collecting all possibilities for the $\alpha \beta$ final states and 
the $\tau$ decays results in Fig.\ \ref{sigvonm}. We denote with 
``hadronic'' the signal coming from final states which 
also have additional hadronic activity from a $\tau$ decay.
We call ``leptonic'' the signals coming from events 
in which two final state leptons are produced 
from which one escapes the identification criteria. Those 
included therefore most channels, namely all of them except the ones 
with hadronic tau decay. 
One can see that muon events have a smaller cross section than the 
$e$ signal, 
coming from the fact that their mixing with the heavy neutrino 
has the biggest constraint. Mass effects play no significant role. 
If the H1 anomaly is indeed explained by heavy Majorana neutrinos, 
one might ask why only muon events are detected. A possible reason for 
that might lie in the following fact: 
The experimental constraint from \obb{} 
on mixing with a heavy Majorana neutrino reads 
\cite{ee1}
\be \label{obbl}
\sin^2 \theta_e \le 5 \cdot 10^{-8} \;  \frac{m_N}{\rm GeV} . 
\ee
Incorporating this in Fig.\ \ref{sigvonm} gives Fig.\ \ref{sigvonm1}. 
Now the electron signal is far below the muon signal. 
In Fig.\ \ref{muonsigvonm} we plot how the cross section 
for the production of a $\mu$ is composed. 
The biggest contribution comes from the $\mu\tau$ channel, which has its 
reason in the mentioned factor $\simeq 3$ 
relative to the $\alpha\alpha$ channels and 
the high hadronic tau branching ratio,  
BR$(\tau \to  \nu_\tau \, \rm{ hadrons}) \simeq \frac{2}{3}$.\\
An additional reason why the $\tau \alpha$ channels add higher 
contributions is that when the 
tau decays it distributes its momentum to three particles, i.e.\ 
the $p_T$ is in general lower and it can therefore escape the 
$p_T \ge 10$ GeV cut more easily and thus has a higher cross section. 
Addmittedly, the process gives only a tiny signal: Multiplying the 
cross section with the 36.5 pb$^{-1}$ luminosity H1 analyzed, gives 
$10^{-8} \ldots 10^{-9}$. However, as explained at the end of the previous 
section, one has to check every possible appearance of Majorana mass 
terms in order to get information about the mass matrix.\\  
One sees that many LSD signals produce the same single lepton signature. 
In the next section we will discuss possibilities to distinguish different 
original final states from the measured ones.

\section{\label{sigobs}Signals and observables}
\begin{figure}[hb]
\setlength{\unitlength}{1cm}
\vspace{0.3cm}
\begin{center}
\epsfig{file=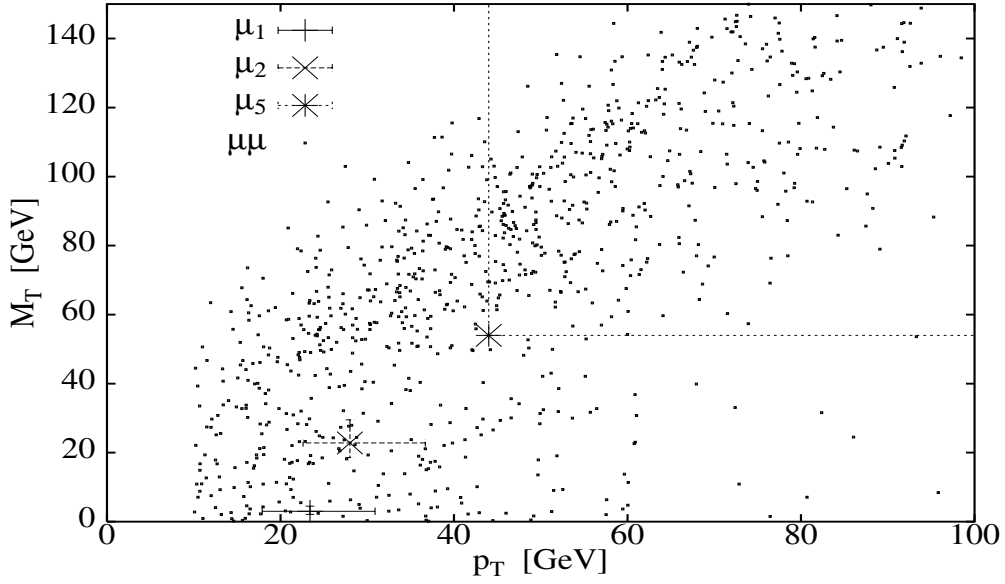,width=13cm,height=8cm}
\end{center}
\caption{\label{mm1}Scatter plot of $p_T$ of the measured lepton 
against $M_T$ for the channel 
$e^+ p \to \aen \mu^+ \mu^+ X$ with one muon escaping the identification 
criteria.} 
\end{figure}

\begin{figure}[hb]
\setlength{\unitlength}{1cm}
\vspace{0.3cm}
\begin{center}
\epsfig{file=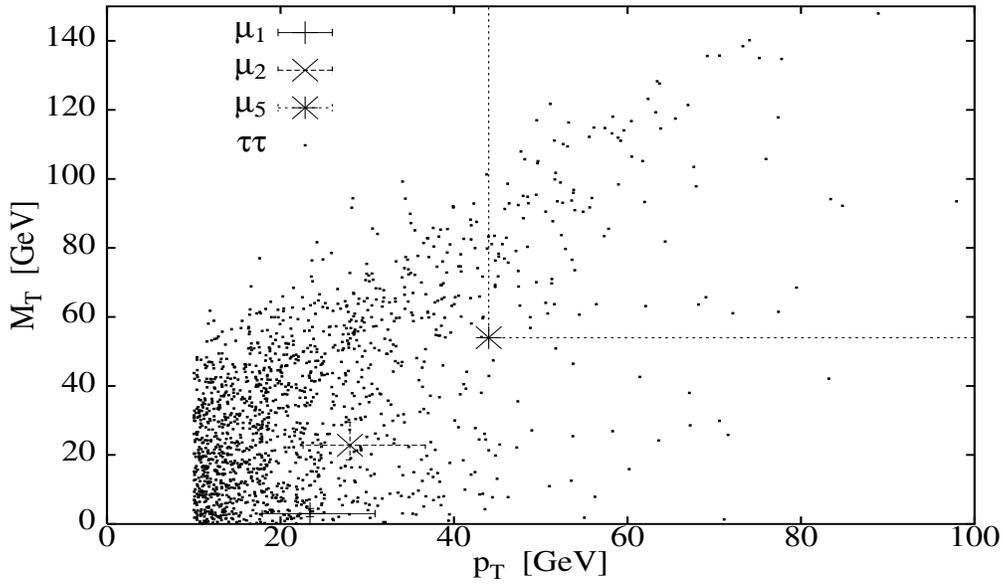,width=13cm,height=8cm}
\end{center}
\caption{\label{tt1a}Scatter plot of $p_T$ of the measured lepton 
against $M_T$ for the channel 
$e^+ p \to \aen \tau^+ \tau^+ X$ with one tau decaying hadronically.} 
\end{figure}

\begin{figure}[hb]
\setlength{\unitlength}{1cm}
\vspace{0.3cm}
\begin{center}
\epsfig{file=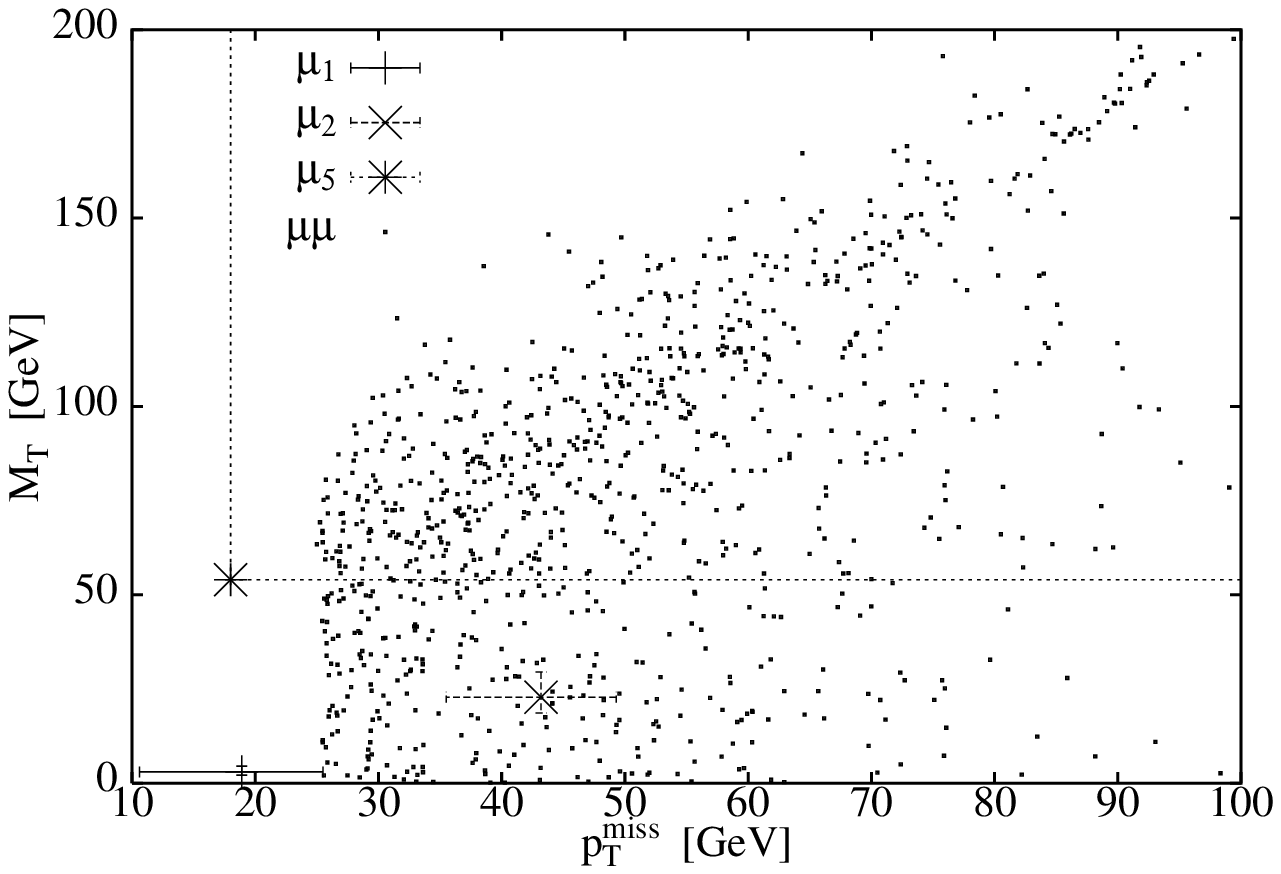,width=13cm,height=8cm}
\end{center}
\caption{\label{mm2}Scatter plot of $\Slash{p}_T$ against $M_T$ 
for the channel $e^+ p \to \aen \mu^+ \mu^+ X$ with one 
muon escaping the identification criteria.} 
\end{figure}

\begin{figure}[hb]
\setlength{\unitlength}{1cm}
\vspace{0.3cm}
\begin{center}
\epsfig{file=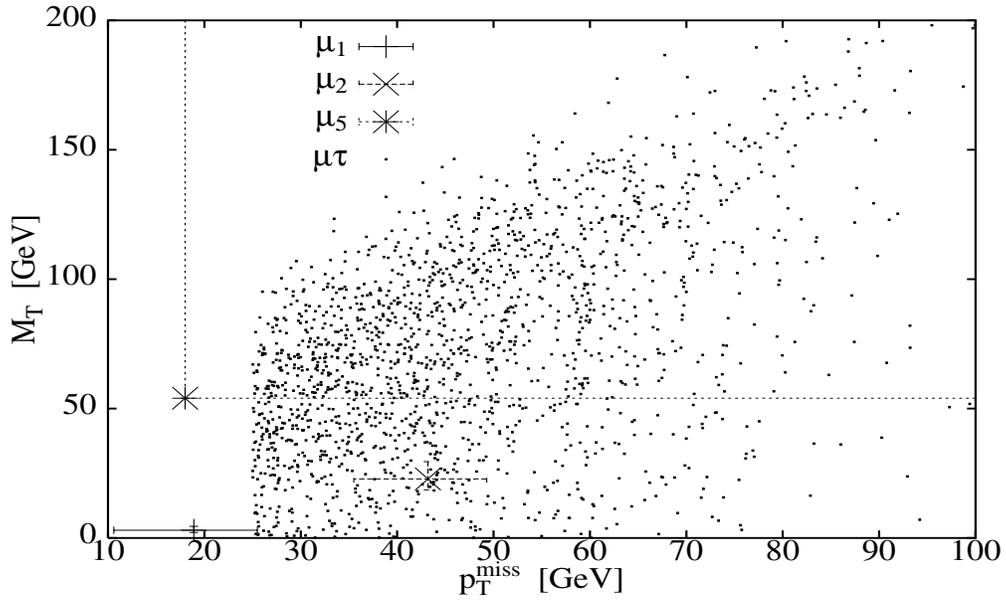,width=13cm,height=8cm}
\end{center}
\caption{\label{mt2a}Scatter plot of $\Slash{p}_T$ 
against $M_T$ for the channel 
$e^+ p \to \aen \tau^+ \mu^+ X$ with the tau decaying hadronically.} 
\end{figure}

Some kinematic quantities of H1's positive muon events are 
given in Table \ref{kintab}. 
In their analysis \cite{isol2} using $36.5 \pm 1.1$ pb$^{-1}$ 
luminosity at $E_p = 820$ GeV and $E_e$ = 27.5 GeV,  
6 events were found 
(0 $e^+$, 1 $e^- $, 2 $\mu^+$, 2 $\mu^-$ and 1 $\mu$ of undetermined charge), 
where about 2 $e$ and 1 $\mu$ are expected from SM processes. 
From those, the most important ones are $W$ production, NC events 
(for $e^+$ events) 
and photon--photon interactions ($\mu^\pm$). We also include the 
event with undetermined charge. The $e^-$ and one $\mu^-$, which also 
has a $e^+$, are very likely to stem from $W$ production. 
In the following we will plot all distributions for $M_N = 200$ GeV since 
the qualitative conclusions we draw remain valid for all masses considered. 
In our analysis it turned out that --- with the kinematics from Table 
\ref{kintab} --- scatter plots of 
$p_T$, $\Slash{p}_T$ and the transverse mass 
$M_T = \sqrt{(\Slash{p}_T + p_T)^2 - (\Slash{\vec{p}}_T + \vec{p}_T)^2}$ 
are most useful. We stress again that for different final states the 
composition of the missing transverse momentum can be made of two 
particles (for $\mu\mu$ or $ee$ final states) to 6 (1 lepton and 
5 neutrinos, $\tau\tau$ channel with two leptonic decays) 
thus changing the area in which events 
populate, say, the $M_T$--$p_T$ space.  
The transverse momentum is also very 
sensitive on the original final state since $\tau$ decays share the  
initial momentum to three particles thereby reducing the average $p_T$.  
This is displayed in Fig.\ \ref{mm1} ($\mu\mu$ final state) 
and Fig.\ \ref{tt1a} ($\tau\tau$, one hadronic decay). If both 
taus decay leptonically the distribution looks similar. 
Obviously, the $\tau\tau$ case has in general low $p_T$ and $M_T$, 
whereas the $\mu\mu$ case displays an uniform distribution with a slight 
band in the center region indicated.\\
Turning now to $\Slash{p}_T$ we see in Figs.\ \ref{mm2} ($\mu\mu$) 
and \ref{mt2a} ($\mu\tau$, hadronic decay) 
that the situation is not as clear in the ``mixed'' channels, i.e.\  
channels with two different final state charged leptons. Though the 
population in $\Slash{p}_T$--$M_T$ space is different (lower values in 
the latter case), it is not as obvious as for the $\mu\mu$/$\tau\tau$ case. 
Due to its low $M_T$, event $\mu_1$ seems to be less probable in all figures, 
though no definite statement can be made.\\
\begin{table}[ht]
\begin{center}
\begin{tabular}{c|c|c|c}
& $\mu_1$ & $\mu_2$ & $\mu_5$ \\ \hline \hline
$p_T^l$ & $ 23.4^{+7.5}_{-5.5}$  
& $ 28.0^{+8.7}_{-5.4} $ & $ > 44 $ \\ \hline
$\Slash{p}_T$ & $  18.9^{+6.6}_{-8.3} $ 
& $ 43.2^{+6.1}_{-7.7} $ & $> 18  $ \\ \hline
$M_T$ & $ 3.0^{+1.5}_{-0.9} $ & $22.8^{+6.7}_{-4.2}  $ & $> 54  $ \\ \hline
$\delta$ & $18.9^{+3.9}_{-3.2}  $ & $17.1^{+2.5}_{-1.7}  $ & $>22  $ \\ \hline
$p_T^X$ & $ 42.2 \pm 3.8 $ & $67.4 \pm 5.4 $ & $30.0 \pm 3.0  $ \\ 
\end{tabular}
\caption{\label{kintab}Kinematical quantities of H1's candidates 
for positively charged muon events. Note that the charge for $\mu_5$ is 
undetermined. 
All values in GeV, taken from \cite{isol2}. The limits 
for $\mu_5$ correspond to 95 $\%$ C.L.}
\end{center}
\end{table}         

\begin{figure}[hb]
\setlength{\unitlength}{1cm}
\vspace{0.3cm}
\begin{center}
\epsfig{file=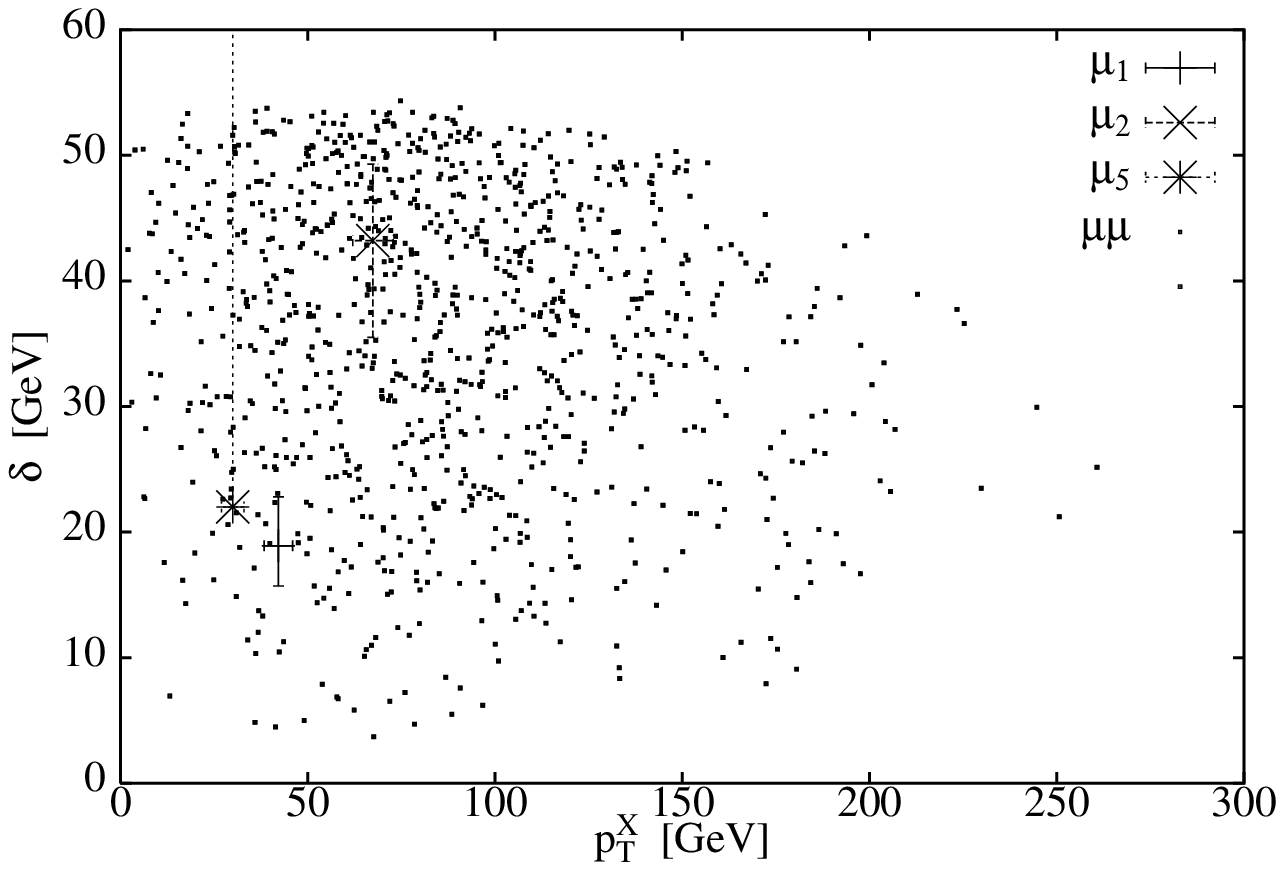,width=13cm,height=8cm}
\end{center}
\caption{\label{hadmm}Scatter plot of $p_T^X$ against $\delta$ 
for the channel $e^+ p \to \aen \mu^+ \mu^+ X$ with one 
muon escaping the identification criteria.} 
\end{figure}

\begin{figure}[hb]
\setlength{\unitlength}{1cm}
\vspace{0.3cm}
\begin{center}
\epsfig{file=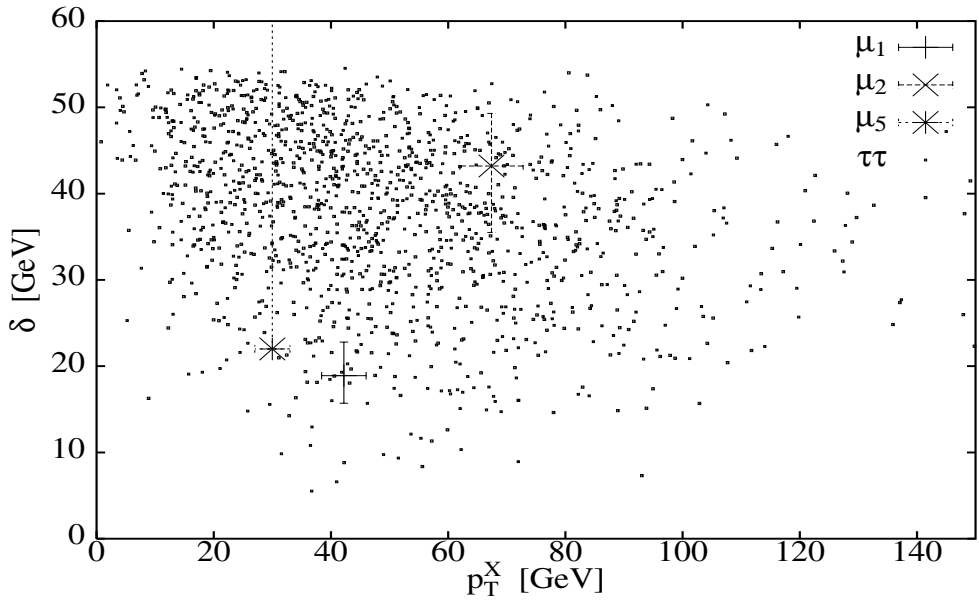,width=13cm,height=8cm}
\end{center}
\caption{\label{hadtt1}Scatter plot of $p_T^X$ against $\delta$ 
for the channel $e^+ p \to \aen \mu^+ \tau^+ X$ with the tau  
decaying hadronically. Note that the $p_T^X$ range is half as wide as 
in the previous figure.} 
\end{figure}

\begin{figure}[hb]
\setlength{\unitlength}{1cm}
\vspace{0.3cm}
\begin{center}
\epsfig{file=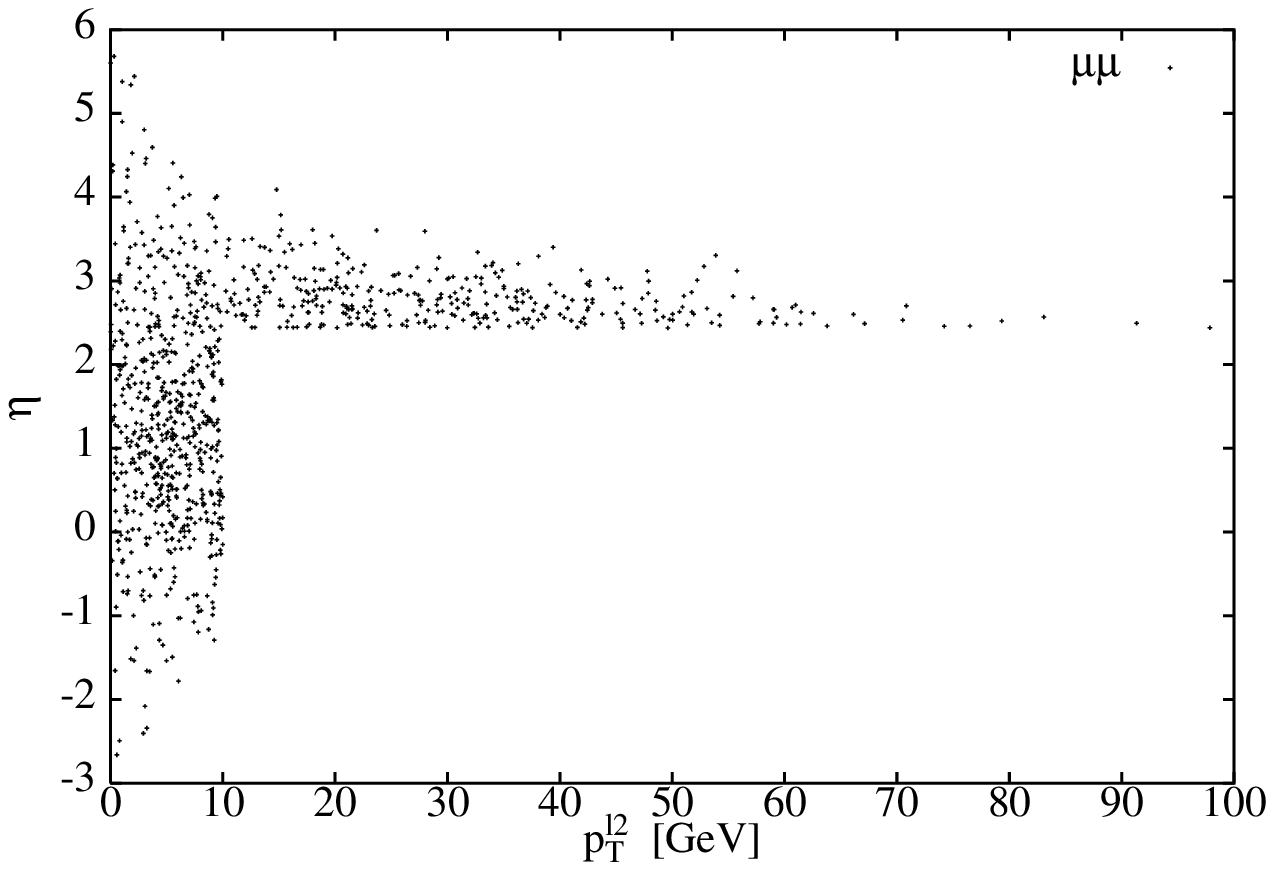,width=13cm,height=8cm}
\end{center}
\caption{\label{lepmm}Scatter plot of $p_T$ of the escaping charged lepton 
against its pseudorapidity for the channel $e^+ p \to \aen \mu^+ \mu^+ X$. 
The sharp edge is due to the applied cuts from H1.} 
\end{figure}

\begin{figure}[hb]
\setlength{\unitlength}{1cm}
\vspace{0.3cm}
\begin{center}
\epsfig{file=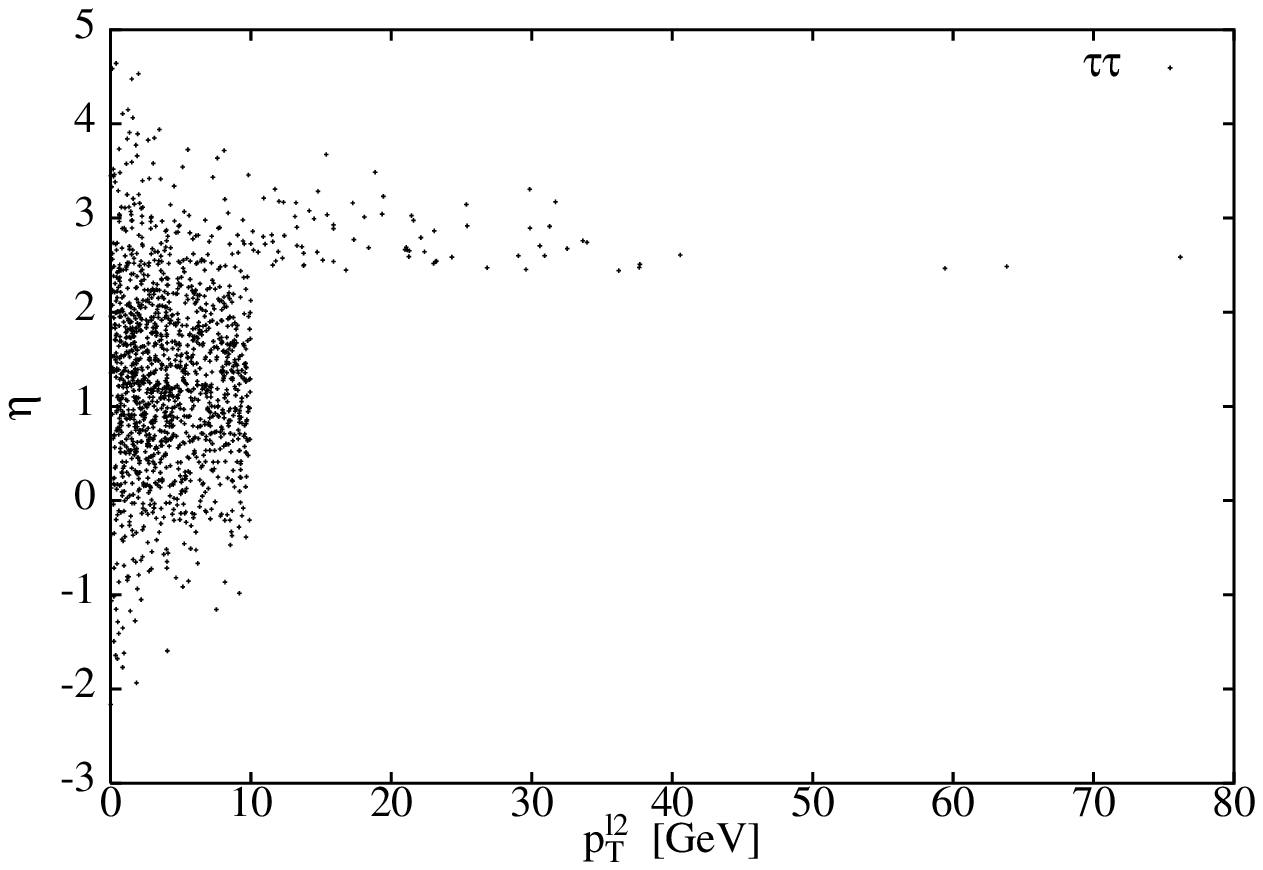,width=13cm,height=8cm}
\end{center}
\caption{\label{leptt}Scatter plot of $p_T$ of the escaping charged lepton 
against its pseudorapidity for the channel $e^+ p \to \aen \tau^+ \tau^+ X$.
The sharp edge is due to the applied cuts from H1.} 
\end{figure}

\begin{figure}[hb]
\setlength{\unitlength}{1cm}
\vspace{0.3cm}
\begin{center}
\epsfig{file=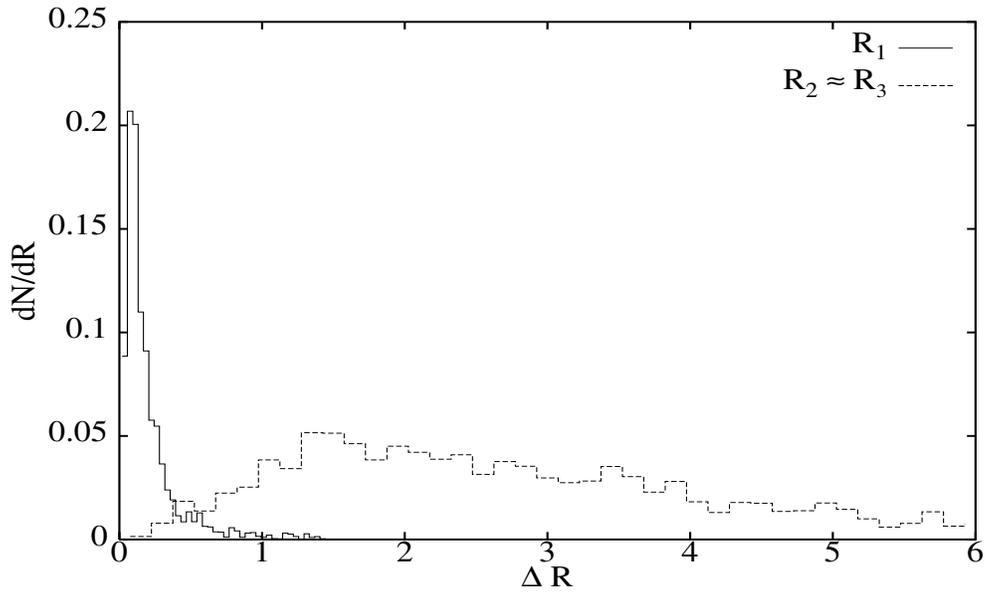,width=13cm,height=8cm}
\end{center}
\caption{\label{jet}Distance of the three jets in $\eta$--$\phi$ space 
with $R_i$ ordered with ascending value for the process 
$e^+ p \rightarrow \overline{\nu}_e \tau^+ \tau^+ X$ with one $\tau$ 
decaying hadronically.} 
\end{figure}

Now we consider the quantities connected with the hadronic remnants. 
We found that the muon signal is composed of roughly 1/3 
purely leptonic final states and 2/3 events with additional jets. 
An interesting quantity is $\delta=\sum E_i ( 1 - \cos \theta_i)$ where 
the sum goes over all measured final state particles. 
In Figs.\ \ref{hadmm} and \ref{hadtt1} one sees that the presence 
of three jets keeps $\delta$ in more or less in the same area whereas 
the transverse momentum of the hadronic system 
$p_T^X$ is shifted towards lower values. 
Here also $\mu_1$ lies in a less crowded area.\\
What are now the signals of the escaping charged lepton (if there is one)? 
In Figs.\ \ref{lepmm} and \ref{leptt} we plot the pseudorapidity $\eta$ 
of the undetected lepton against its transverse momentum. Again, 
the $\mu\mu$ and the $\tau\tau$ case can be distinguished since 
the latter has far lower $p_T$. 
The $\tau$ boosts its decay products more or less in its 
forward direction so that $\eta$ does not alter much. 
The problem one might encounter is that the escaping lepton is 
hiding in the hadronic jet. Demanding a distance of $\Delta R \ge 1.5$ 
reduces the cross section by 10 to 15 $\%$ but does not change the 
distributions in Figs.\ \ref{lepmm} and \ref{leptt}.\\
We mimicked the hadronic $\tau$ decay via two quark jets and 
ignored effects of 
modes like $\tau \to \pi$'s $\nu_\tau$. 
Thus, due to the boost of the $\tau$,  
two of the three jets of events with a hadronically 
decaying tau will be very close together. Fig.\ \ref{jet} displays 
the normalized distribution of the distance in $\eta$--$\phi$ space. 
We denote with $R_i$ the distance ordered with ascending value. 
One distance is centered significantly below one, 
therefore, probably two jets instead of three will be measured. 
However, the $\tau$ identification is hard to do and 
Fig.\ \ref{jet} serves only as an indication of how things might work. 
Information on the jet multiplicity is not given in Ref.\ \cite{isol2}, 
though $\mu_1$ and $\mu_5$ seem to have additional 
separated tracks in their event 
displays as can be seen in Fig.\ 2 of Ref.\ \cite{isol2}.

\section{\label{concl}Conclusions}
In the light of recent data we did a full analysis of the 
analogue of the \onbb{} graph with 
all possible two charged leptons in the final state. 
One lepton can escape the 
identification criteria by either having low transverse momentum 
and/or high pseudorapidity or (if it is a $\tau$) via hadronic decay. 
Signatures of these events are discussed and compared to H1's isolated 
leptons with large missing transverse momentum. 
All their positive muon events lie in the regions typically populated by 
the ``double beta'' process, 
though $\mu_1$ is always in a less crowded area. Due to its 
high errors $\mu_5$ is mostly in the favored region (but for the 
same reason always in the region populated by SM processes \cite{isol2}). 
Though the process represents an attractive explanation, the smallness 
of the expected signal might spoil our interpretation. Nevertheless, 
any information about mass matrix entries and Majorana particles 
is very important and worth 
looking for, regardless of the small expected signal. 
Other extensions of the SM might give larger signals to the discussed 
final states 
and it is then helpful to know how the 
``SM + heavy Majorana'' extension contributes.\\  
To confirm our hypothesis direct production of heavy Majorana 
neutrinos will be the only possibility since cross sections or decay widths 
of other \obb--like processes are probably too small to be 
detected \cite{ich}. 
Here, either HERA itself or LHC will be the 
candidates for this observation. 
We did not consider the mass reconstruction of the Majorana since 
the large number of unmeasured particles does not permit that.\\ 
Other proposed explanations \cite{isol1} for the events 
were FCNC interactions 
(topologically identical to leptoquark production) 
or high $p_T$ jets from which one fakes a muon signal. 
Production of supersymmetric particles 
is suggested in \cite{kon} to 
explain the results. 
A definite answer regarding all detector/identification issues 
can only be given by the collaboration itself. 
From the ``new physics'' side we believe that 
massive neutrinos provide one of the most natural possibilities. 

\hspace{3cm}
\begin{center}
{\bf \large Acknowledgments}
\end{center}
Its a pleasure to thank M.\ Flanz for helpful discussions and careful 
reading of the manuscript. 
This work has been supported in part (W.\ R.) by the
``Bundesministerium f\"ur Bildung, Wissenschaft, Forschung und
Technologie'', Bonn under contract No. 05HT9PEA5. 
Financial support from the Graduate College
``Erzeugung und Zerf$\ddot{\rm a}$lle von Elementarteilchen''
at Dortmund university (W.\ R.) is gratefully acknowledged.

\newpage


\newpage

\newpage

\newpage

\newpage

\newpage

\newpage

\end{document}